\documentclass[12pt,a4paper]{article}

\usepackage{graphicx,subfigure}
\usepackage{amsmath,amssymb}
\usepackage{color}

\title{Oscillations of the number of immune system cells in a space-velocity thermostatted kinetic theory model of tumor growth} 

\author{L\'eon Masurel$^1$, Carlo Bianca$^2$, and Annie Lemarchand*$^1$\\
$^{1}$ {\small {\it Sorbonne Universit\'e, Centre National de la Recherche Scientifique CNRS,}}\\
       {\small {\it Laboratoire de Physique Th\'eorique de la Mati\`ere Condens\'ee, LPTMC,}}\\
       {\small {\it 4 place Jussieu, case courrier 121, 75252 Paris Cedex 05, France}}\\
$^{2}$ {\small {\it \'Ecole Sup\'erieure d'Ing\'enieurs en G\'enie \'Electrique, Productique et Management Industriel,}}\\
       {\small {\it Laboratoire Quartz EA 7393,}}\\ 
       {\small {\it Laboratoire de Recherche en Eco-innovation Industrielle et Energ\'etique,}}\\
       {\small {\it 13 Boulevard de l'Hautil, 95092 Cergy Pontoise Cedex, France}}
}

\begin{document}

\maketitle

\begin{abstract}
The competition between cancer cells and immune system cells in inhomogeneous conditions is described at cell scale within the framework of the thermostatted kinetic theory.
Cell learning is reproduced by increased cell activity during favorable interactions. The cell activity fluctuations are controlled by a thermostat.
The direction of cell velocity is changed according to stochastic rules mimicking a dense fluid. 
We develop a kinetic Monte Carlo algorithm inspired from the direct simulation Monte Carlo (DSMC) method initially used for dilute gases. The simulations generate
stochastic trajectories sampling the kinetic equations for the distributions of the different cell types. The 
evolution of an initially localized tumor is analyzed.
Qualitatively different behaviors are observed as the field regulating activity fluctuations decreases.
For high field values, i.e. efficient thermalization, cancer is controlled. 
For small field values, cancer rapidly and monotonously escapes from immunosurveillance. 
For the critical field value separating these two domains, the 3E's of immunotherapy are reproduced, 
with an apparent initial elimination of cancer, a long quasi-equilibrium period followed by large fluctuations, 
and the final escape of cancer, even for a favored production of immune system cells.  
For field values slightly smaller than the critical value, 
more regular oscillations of the number of immune system cells are spontaneously observed in agreement with clinical observations.
The antagonistic effects that the stimulation of the immune system may have on oncogenesis are reproduced in the model
by activity-weighted rate constants for the autocatalytic productions of immune system cells and cancer cells. 
Local favorable conditions for the launching of the oscillations are met in the fluctuating inhomogeneous system, able to generate a small cluster of immune system cells with larger activities than those of the surrounding cancer cells.
\end{abstract}

\section{Introduction}
A large variety of models of interactions between agents have been developed within the framework of thermostatted kinetic theory \cite{wondmagegne,carloPLR2012} and notably applied to traffic \cite{coscia2011}, swarming \cite{carrillo} and cell competition \cite{carlo2,carlo3,jcp,physicaa}. 
The interactions between cancer cells and immune system cells are known to be complex. Stimulating the immune system of a patient can lead to disappointing results and conversely, chemotherapy may boost immune system response \cite{zitvogel}. 
C-Reactive Protein (CRP) levels have been used as markers of inflammation in response to antigen exposure in relation to the progression of many cancers, such as melanoma and lung cancer \cite{mahmoud}. Oscillations of CRP concentration in advanced cancer patients have been observed and interpreted in terms of immune regulatory cycles \cite{coventry,sawamura}. The oscillations of CRP concentration reveal the antagonistic effects of immune system stimulation on a tumor. 
Many macroscopic models of interactions between cancer and immune system leading to oscillating solutions for concentrations can be found in the recent
literature \cite{lejeune08,onofrio10,liu12,wilkie13,bi14,wenbo17,dritschel18,alvarez19} but without direct connection with a description at cell scale.
We recently designed a minimal model of interactions between cancer and immune system cells in a homogeneous system, ignoring the spatial description of a tumor \cite{jcp,physicaa}. 
In this work we introduce a space-velocity description of tumor growth at cell scale in the framework of thermostatted kinetic theory. 
The model includes autocatalytic cell production and increased cell activity during favorable interactions. The fluctuations of cell activity are regulated by the so-called thermostat. Jump processes are introduced to randomize the direction of cell velocities.
We design a kinetic Monte Carlo algorithm inspired by the direct simulation Monte Carlo (DSMC) method initially devoted to dilute gases \cite{bird,garcia} and simulate the evolution of cell position, velocity, nature, and activity. Our aim is to check if the inhomogeneous model contains enough biological ingredients to reproduce the complex antagonistic effects observed in cancer evolution, including the three Es (elimination, equilibrium, and escape) \cite{dunn2004,alvarez19} of immunotherapy and spontaneous oscillations in the number of immune system cells during tumor growth.

The paper is organized as follows. In section 2, we present the inhomogeneous model and the kinetic equations governing the evolution of the probability distributions of cancer, immune system and normal cells. The kinetic Monte Carlo algorithm is made precise. Simulation results for an initially localized tumor are presented in section 3. Specifically, we examine if, for a poorly effective thermalization, the number of cancer cells rapidly drops, then slightly increases during a long quasi-steady regime, and eventually explodes, although the autocatalytic production of immune system cells is much more favored than cancer cell formation. The existence of oscillations for the total number of immune system cells is also discussed. A sensitivity analysis is performed and the different behaviors observed as the field that regulates activity and the rate constant related to the production of immune system cells vary are shown. The variations of the pseudo-period of oscillations with the field and the rate constant are determined. Section 4 contains conclusions.

\section{The inhomogeneous thermostatted kinetic theory model}
The competition between immune system cells and cancer cells can be compared to the interactions between immune system cells and viruses. In both cases, the foreign cells induce responses from the immune system \cite{blankenstein,vesely,levy}. Different types of cells are involved in the defense mechanism \cite{siegrist,leo}.
Dentritic cells ingest cancer cells, isolate antigens and present them to T cells, leading to their activation and proliferation \cite{guermonprez}. T-effector cells are able to destroy cancer cells and T-regulatory cells control the number of lymphocytes and prevent their overproduction \cite{zitvogel,gonzalez}. In parallel with the learning of the immune system cells, the learning of the cancer cells takes place, which may lead to cancer escape. Through cell interactions, the cancer cells learn not to be detected, misleading the T-regulatory cells which in turn limit the production of immune system cells and let cancer cells proliferate \cite{dunn2002,igney}.\\

Recently, we have introduced a homogeneous model of competition between cancer cells and immune system cells which reproduces cell interactions, activation, and learning processes \cite{jcp,physicaa} and accounts for the three Es of immunotherapy.
We propose to develop a space-velocity description of tumor growth, which requires to model the updating of cell position $x$ and velocity ${\bf v}$ in addition to their nature $j$ and activity $u$.
As in the homogeneous model, we introduce normal cells $n$, cancer cells $c$, and the immune system cells $i$ as a whole with the aim of introducing as few variables and parameters as possible and providing qualitative interpretations of clinical observations.

Inhomogeneous initial conditions with tumor cells located in the middle of the system are preferably considered.
A Gaussian distribution $P_u(u)$ of mean $\mu$ and variance $\sigma$ is used to generate the initial activities of the cells, regardless of their nature. Possible heterogeneity in initial activity distribution may be considered. All the cells are supposed to have the same speed which remains constant. The direction of cell velocity ${\bf v}$ is defined by an angle $\theta_v \in ]0,2\pi]$ randomly chosen using a uniform distribution $P_{\bf v}({\theta_v})$.

\subsection{The different processes affecting cell evolution}
Cell nature $j=c,i,n$ and activity $u$ may change during local interactions that are assumed to leave position $x$ and velocity ${\bf v}$ of cancer cells and immune system cells unchanged.
We consider only three processes which include interaction, activation, proliferation or death, in a similar manner as autocatalytic chemical processes
\begin{eqnarray}
\left\{ 
\begin{array}{ll}
\label{cn}
c({\bf v},u)+n({\bf v'},u')  \begin{array}{c} 
\mbox{\footnotesize $\kappa_{cn}({\bf v},u,{\bf v'},u')$} \\ \longrightarrow  \\ \mbox{}
\end{array}  c({\bf v},u+\epsilon)+c({\bf v'},u')\\
R  \longrightarrow  n({\bf v''},u'')\\ 
\end{array}
\right. \\
\label{ic}
i({\bf v},u)+c({\bf v'},u') \begin{array}{c} 
\mbox{\footnotesize $\kappa_{ic}({\bf v},u,{\bf v'},u')$} \\ \longrightarrow \\ \mbox{}
\end{array} i({\bf v},u+\epsilon)+i({\bf v'},u')\\ 
\label{ci}
c({\bf v},u)+i({\bf v'},u')  \begin{array}{c} 
\mbox{\footnotesize $\kappa_{ci}({\bf v},u,{\bf v'},u')$} \\ \longrightarrow \\ \mbox{}
\end{array} c({\bf v},u+\epsilon)+c({\bf v'},u')
\end{eqnarray}
where the rate constants $\kappa_{cn}({\bf v},u,{\bf v'},u')$, $\kappa_{ic}({\bf v},u,{\bf v'},u')$, and $\kappa_{ci}({\bf v},u,{\bf v'},u')$ have the following nontrivial dependence on the activities and the velocities of the interacting couple:
\begin{eqnarray}
\label{kappacj}
\kappa_{cj}({\bf v},u,{\bf v'},u')&=&k_{cj}(u-u')H(u-u')H\left(\theta_{\rm inter} -  \theta({\bf v},{\bf v'}) \right) \, {\rm for} \, j=n,i\\
\label{kappaic}
\kappa_{ic}({\bf v},u,{\bf v'},u')&=&k_{ic}(u-u')H(u-u')H\left(\theta_{\rm inter} - \theta({\bf v},{\bf v'}) \right)
\end{eqnarray}
where $k_{cn}$, $k_{ic}$ and $k_{ci}$ are constant, $\theta_{\rm inter} \in ]0,\pi]$ is the maximum angle between the velocity directions permitting interaction and
$\theta({\bf v},{\bf v'})$ is defined as follows:
\begin{eqnarray}
\theta({\bf v},{\bf v'})=\left\{\begin{array}{ll}
\mid \theta_v - \theta_{v'} \mid \, {\rm if} \, \mid \theta_v - \theta_{v'} \mid \leq \pi \\
\mid \mid \theta_v - \theta_{v'} \mid -2\pi \mid \, {\rm if} \, \mid \theta_v - \theta_{v'} \mid > \pi  
 \end{array} \right.                                          
\end{eqnarray}
where $\theta_v \in ]0,2\pi]$ is the angle defining the direction of ${\bf v}$. 
The condition on the directions of the velocities of the interacting pair imposed by the Heaviside function 
$H\left(\theta_{\rm inter} - \theta({\bf v},{\bf v'}) \right)$ ensures that the two cells remain close during a sufficient time for activation and learning to take place.\\

The learning process is reproduced by the small increase $\epsilon$ of the activity of the cell with the already larger activity before the interaction. 
The process described in Eq.~(\ref{cn}), which has the form of an autocatalytic production of cancer cells, accounts for both the mutation of a normal cell into a cancer cell and the division of a cancer cell. The specific rate constant expression ensures that
proliferation of cancer cells is more favorable when they have already learned to blend into their environment, i.e. have increased their activity $u$. 
The reservoir $R$ of normal cells maintains their number 
constant and plays also the role of reservoir of activity.
Indeed, the condition on the activities of the interacting cells imposed by the Heaviside function $H(u-u')$ introduces a selection.
The normal cell which disappears has a smaller activity than the interacting cancer cell, which could induce a non physical increase in the mean activity of the remaining normal cells. The model provides for reinjecting a normal cell with a randomly chosen activity $u''$ using the initial Gaussian distribution $P_u(u)$ of fixed mean value $\mu$, so that the mean activity of normal cells remains constant, consistent with biology. The velocity ${\bf v''}$ of the reinjected normal cell is randomly chosen using
the initial uniform distribution $P_{\bf v}({\theta_v})$.

The process given in Eq.~(\ref{ic}) corresponds to the autocatalytic production of immune system cells and accounts for both the death of a cancer cell attacked by a T-effector cell and the proliferation of immune system cells with increased activity, i.e. the knowledge of the presence of cancer cells in their environment. As in Eq.~(\ref{cn}), the rate constant depends on a Heaviside step function $H(u-u')$ implying that the increase of the number of immune cells is locally more probable when they have a larger activity than the neighboring cancer cells.

The regulation of the number of immune cells by T-regulatory cells and the learning process of cancer cells are taken into account in the third process given in Eq.~(\ref{ci}). This process accounts for both the death of an immune system cell and the division of a cancer cell. The rate constant being proportional to $H(u-u')$, the proliferation of cancer cells locally occurs if their activity is larger than the one of the surrounding immune system cells. Hence, the model accounts for the ability of trained cancer cells to mislead the immune system and trigger the detrimental response of T-regulatory cells.

In order to introduce a regulation of the explosive cell production due to autocatalytic processes, we impose that the rate of each process is proportional to the difference of activities of the interacting pair. Specifically, according to Eq. (\ref{ci}), further growth of a cluster of cancer cells with a smaller activity than the surrounding immune system cells is not likely to be observed. Following Eq. (\ref{ic}), an analogous phenomenon is expected for immune system cells: A local increase in the number of immune system cells will not be sustained if their activity is smaller that the one of the neighboring cancer cells. 
In order to mimic the preferential increase in the number of cancer cells due to the deleterious response of T-regulatory cells rather than by mutation of normal cells, we assign values of $k_{ci}$ in Eq. (\ref{ci}) larger than the values of $k_{cn}$ in Eq. (\ref{cn}).\\

The processes given in Eqs. (\ref{cn}-\ref{ci}) associate learning process with systematic increased cell activity. However, cell interactions that are not explicitly taken into account in the scheme make learning less effective. In the model the regulation of the activity fluctuations is ensured by a thermostat mimicking information loss due to dissipation. Thermalization, characterized by a field $E$ and a friction coefficient $\alpha$, acts on the activity $u_k$ of each cell $k$ according to \cite{wondmagegne,carlo2}
\begin{equation}
\label{newton}
\frac{{\rm d}u_k}{\rm{d}t}=E-\alpha u_k
\end{equation}
Choosing
\begin{equation}
\label{alpha}
\alpha=\frac{\langle u \rangle E}{\langle u^2 \rangle}
\end{equation}
where $\langle u \rangle$ is the mean activity of the entire system, ensures that the second moment $\langle u^2 \rangle$ remains constant. \\
  
All cells are supposed to move at constant speed in all directions. We introduce a stochastic turning operator that modifies the direction of cell velocities inside a cone of apex angle $\theta_{\rm jump}$ with $0 < \theta_{\rm jump} \leq \pi/2$.
Hence, the rate constant associated with the velocity jumps obeys
\begin{equation}
\label{kappav}
\kappa_v({\bf v},{\bf v'}) = k_v H\left(\theta_{\rm jump} - \mid \theta_{\bf v}-\theta_{\bf v'} \mid \right)
\end{equation}
where $k_v$ is constant.

\subsection{The kinetic equations}
The distribution function $f_j$ for cells of type $j=c,i,n$ obeys the following kinetic equation
\begin{equation}
\label{kineqj}
\left(\partial_{t} + {\bf v} \cdot \nabla_{\bf x} \right) f_j(t,{\bf x},{\bf v},u)+ \partial_{u} \left((E-\alpha u)f_j(t,{\bf x},{\bf v},u) \right) = I_j + V_j
\end{equation}
where ${\bf v} \cdot \nabla_{\bf x}$ is the standard advection operator, $\partial_{u} \left((E-\alpha u)f_j \right)$ is the thermalization operator, $I_j$ is the interaction operator and $V_j$ is the velocity randomization operator.
Following Eqs. (\ref{cn}-\ref{ci}), the interaction operator $I_c$ associated with cancer cells reads
\begin{eqnarray}
\label{Ic}
I_c &=& \int_{\mathbf{R}^3} \int_{\mathbf{R}^+} \kappa_{cn}({\bf v},u-\epsilon,{\bf v'},u') f_c(t,{\bf x},{\bf v},u-\epsilon) f_n(t,{\bf x},{\bf v'},u') {\rm d}{\bf v'} {\rm d} u' \nonumber \\
    &+& \int_{\mathbf{R}^3} \int_{\mathbf{R}^+}\kappa_{cn}({\bf v'},u',{\bf v},u) f_c(t,{\bf x},{\bf v'},u') f_n(t,{\bf x},{\bf v},u) {\rm d}{\bf v'} {\rm d}u' \nonumber\\ 
    &-& \int_{\mathbf{R}^3} \int_{\mathbf{R}^+} \kappa_{cn}({\bf v},u,{\bf v'},u') f_c(t,{\bf x},{\bf v},u) f_n(t,{\bf x},{\bf v'},u') {\rm d}{\bf v'} {\rm d} u' \nonumber \\
    &-& \int_{\mathbf{R}^3} \int_{\mathbf{R}^+} \kappa_{ic}({\bf v'},u',{\bf v},u) f_i(t,{\bf x},{\bf v'},u')f_c(t,{\bf x},{\bf v},u) {\rm d}{\bf v'} {\rm d}u' \nonumber\\
    &+& \int_{\mathbf{R}^3} \int_{\mathbf{R}^+}\kappa_{ci}({\bf v},u-\epsilon,{\bf v'},u')f_c(t,{\bf x},{\bf v},u-\epsilon) f_i(t,{\bf x},{\bf v'},u') {\rm d}{\bf v'} {\rm d}u' \nonumber\\
    &+& \int_{\mathbf{R}^3} \int_{\mathbf{R}^+} \kappa_{ci}({\bf v'},u',{\bf v},u) f_c(t,{\bf x},{\bf v'},u') f_i(t,{\bf x},{\bf v},u) {\rm d}{\bf v'} {\rm d}u' \nonumber\\
    &-& \int_{\mathbf{R}^3} \int_{\mathbf{R}^+}\kappa_{ci}({\bf v},u,{\bf v'},u')f_c(t,{\bf x},{\bf v},u) f_i(t,{\bf x},{\bf v'},u') {\rm d}{\bf v'} {\rm d}u' 
\end{eqnarray}
with $\kappa_{cj}({\bf v},u,{\bf v'},u')$ and $\kappa_{ic}({\bf v},u,{\bf v'},u')$ given in Eqs. (\ref{kappacj}) and (\ref{kappaic}).
The two first terms of the right-hand side account for the positive contributions to the distribution function of cancer cells of velocity ${\bf v}$ and activity $u$ after the interaction of a cancer cell and a normal cell through the process given in Eq. (\ref{cn}). Specifically,
the first term corresponds to the formation of a cancer cell with the desired properties from a cancer cell of initial activity $u-\epsilon$ 
after update by the increment $\epsilon$. All contributions are obtained by integrating 
over all acceptable velocities ${\bf v'}$ and activities $u'$ of the interacting normal cell. 
The second term is obtained when assigning the velocity ${\bf v}$ and activity $u$ of the disappearing normal cell to a newly formed cancer cell 
and integrating over all acceptable velocities ${\bf v'}$ and activities $u'$ of the interacting cancer cell.
The fourth term is related to the process given in Eq. (\ref{ic}) and the fifth to seventh terms are associated with Eq. (\ref{ci}).
The interaction operator $I_i$ for immune system cells is formed following analogous rules as for cancer cells:
\begin{eqnarray}
\label{Ii}
I_i &=& \int_{\mathbf{R}^3}\int_{\mathbf{R}^+}\kappa_{ic}({\bf v},u-\epsilon,{\bf v'},u') f_i(t,{\bf x},{\bf v},u-\epsilon)f_c(t,{\bf x},{\bf v'},u'){\rm d}{\bf v'} {\rm d}u' \nonumber \\
    &+& \int_{\mathbf{R}^3}\int_{\mathbf{R}^+} \kappa_{ic}({\bf v'},u',{\bf v},u) f_i(t,{\bf x},{\bf v'},u')f_c(t,{\bf x},{\bf v},u) {\rm d}{\bf v'} {\rm d}u' \nonumber\\
    &-& \int_{\mathbf{R}^3}\int_{\mathbf{R}^+}\kappa_{ic}({\bf v},u,{\bf v'},u') f_i(t,{\bf x},{\bf v},u)f_c(t,{\bf x},{\bf v'},u'){\rm d}{\bf v'} {\rm d}u' \nonumber \\
    &-& \int_{\mathbf{R}^3}\int_{\mathbf{R}^+} \kappa_{ci}({\bf v'},u',{\bf v},u) f_c(t,{\bf x},{\bf v'},u') f_i(t,{\bf x},{\bf v},u) {\rm d}{\bf v'} {\rm d}u'
\end{eqnarray}
For normal cells, the interaction operator $I_n$ reads
\begin{eqnarray}
\label{In}
I_n &=& -\int_{\mathbf{R}^3}\int_{\mathbf{R}^+} \kappa_{cn}({\bf v'},u',{\bf v},u) f_c(t,{\bf x},{\bf v'},u') f_n(t,{\bf x},{\bf v},u) {\rm d}{\bf v'} {\rm d}u' \\
&+& P_u(u) P_{\bf v}({\theta_v})\iint_{\mathbf{R}^3}\iint_{\mathbf{R}^+}  \kappa_{cn}({\bf v'},u',{\bf v''},u'')f_c(t,{\bf x},{\bf v'},u') f_n(t,{\bf x},{\bf v''},u'') {\rm d}{\bf v'} {\rm d}{\bf v''} {\rm d}u' {\rm d}u'' \nonumber
\end{eqnarray}
The second term of the right-hand side is related to the action of the reservoir, which injects a normal cell 
of velocity ${\bf v}$ and activity $u$ exactly 
at the same rate as a normal cell of velocity ${\bf v''}$ and activity $u''$ just disappeared through Eq. (\ref{cn}). The contribution to the evolution of $f_n(t,{\bf x},{\bf v},u)$
includes integration over all velocities ${\bf v''}$ and weighting by $P_{\bf v}({\theta_v})$, integration over all activities $u''$ and weighting by $P_u(u)$, as well as
integration over all acceptable velocities ${\bf v'}$ and activities $u'$ of the interacting cancer cell.\\

Finally, the velocity randomization operator $V_j$ which models the velocity-jump process inside a cone reads
\begin{eqnarray}
\label{Vj}
V_j &=& \int_{\mathbf{R}^3} \kappa_v({\bf v},{\bf v'}) \left( f_j(t,{\bf x},{\bf v'},u)-f_j(t,{\bf x},{\bf v},u) \right){\rm d}{\bf v'} 
\end{eqnarray}
where the rate constant $\kappa_v({\bf v},{\bf v'})$ of clockwise or anticlockwise jump of velocity direction is given in Eq. (\ref{kappav}).\\

\subsection{The simulation algorithm}
Graeme Bird \cite{bird,garcia} introduced an efficient algorithm, the direct simulation Monte Carlo (DSMC) method,  
to simulate the Boltzmann equations associated with a dilute gas \cite{turingpiotr} and successfully adapted to concentrated solutions \cite{turinggab}.
The performance of DSMC with respect to molecular dynamics is related to considering collisions only for particles belonging to a same spatial box
and following an acceptance-rejection technique to treat the collisions.
Our aim is to show that these hypotheses can be satisfactorily extended to cellular dynamics in order to directly solve the three kinetic equations given in Eq.~(\ref{kineqj}) for the different cell types $j=c,i,n$. We propose the following algorithm.
During a time step $\Delta t$, cell natures, activities, positions and velocities are updated according to the model of interactions, thermalization and velocity jumps
described in the previous section. Specifically, for given values of the constants $k_{cn}$, $k_{ic}$, and $k_{ci}$, the maximum number of interactions between pairs is determined in each spatial box and the corresponding interactions between randomly chosen pairs are actually accepted provided that the conditions on activities and velocities expressed in Eqs. (\ref{kappacj}) and (\ref{kappaic}) are met. The nature and the activity of the interacting cells are then updated. After the interactions in all spatial boxes have been considered, the activity of each cell is thermalized according to Eqs.~(\ref{newton}).
Cell positions are modified according to their velocities, which may result in changing of spatial box.
Finally, the directions of cell velocities are updated following the stochastic deflection process of rate constant given in Eq.~(\ref{kappav}).\\

In the next section we present the simulation results obtained for a two-dimensional inhomogeneous system.

\section{Results}
Two-dimensional simulations are performed in a square of side $l_x=l_y=30$ divided in boxes of side $\Delta l=1$. Periodic boundary conditions are applied. Initially, $20$ normal cells and $20$ immune system cells per box are homogeneously spread in the entire system. 
In order to study the growth of an initially localized tumor, we start from $20$ cancer cells per box in a central discretized disc of diameter $D=5$ and no cancer cells elsewhere. The effects of the rate constant values on the system behavior have been studied in detail in a homogeneous system \cite{jcp,physicaa} and we have checked that
they are not sensitively different in the inhomogeneous case. 
We choose to present the results obtained in a boosted immune system such that $k_{ic} > k_{ci}$. This choice results in the more favorable autocatalytic formation of immune system cells than cancer cells for populations with similar activities.
We impose $k_{ci} >  k_{cn}$ in order to comply with the more probable division of cancer cells under the pressure exerted by regulatory T cells
through Eq. (\ref{ci}) than the mutation of normal cells into cancer cells reproduced by Eq. (\ref{cn}). In order to speed up the simulation, we accept the 
interaction between two cells without condition on the direction of their velocities, which amounts to choosing $\theta_{\rm inter}=\pi$.
The rate constant for velocity deflection $k_v$ and the maximum angle of velocity jump $\theta_{\rm jump}$ are chosen sufficiently large for the randomization of velocity direction to be efficient. We have $k_v=0.1$ and $\theta_{\rm jump}=\pi/6$.\\

Before performing a sensitivity analysis and showing the different behaviors encountered 
as essential parameters of the model vary, we focus on two nontrivial
behaviors reproducing clinical observations, the three Es of immunotherapy \cite{dunn2004} and oscillations of the number of immune system cells \cite{coventry,sawamura}.

\begin{figure}[htb]
\begin{center}
\subfigure[]{\includegraphics[height=7cm]{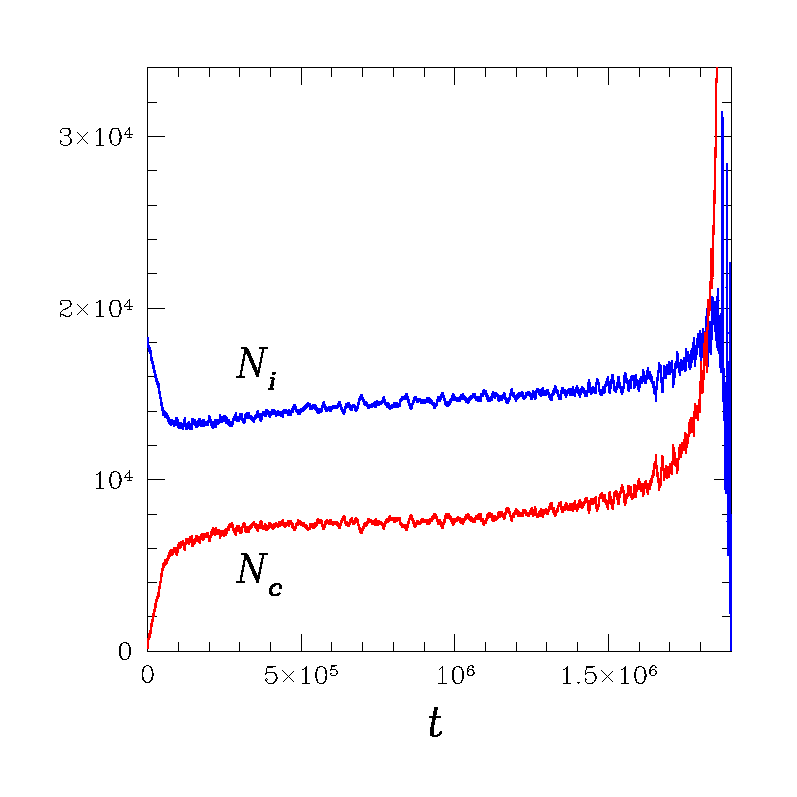}}
\subfigure[]{\includegraphics[height=7cm]{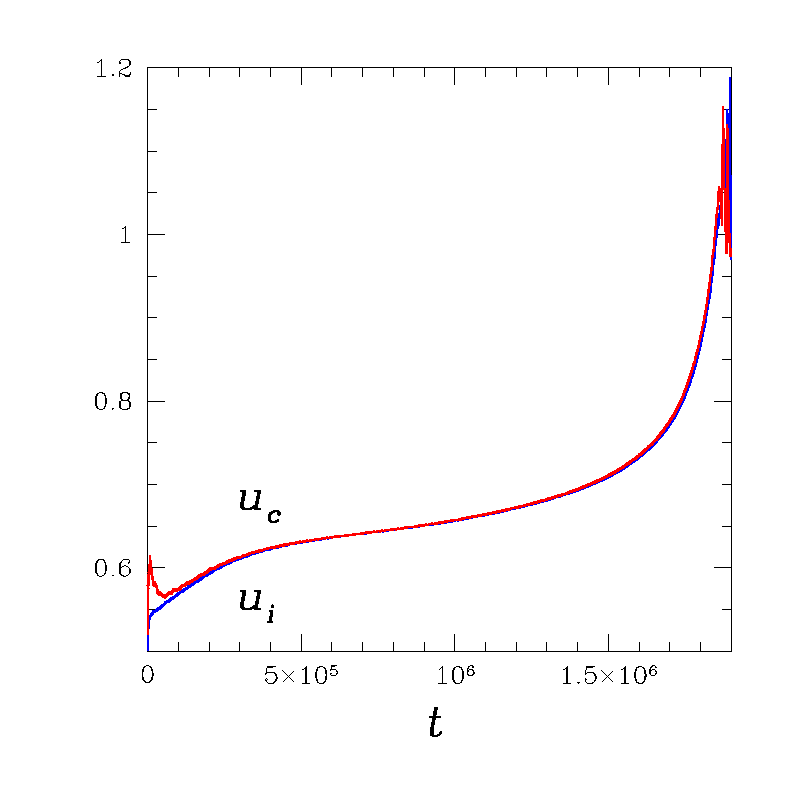}}
\caption{The three Es of immunotherapy.
(a) Evolution of total number $N_i$ of immune system cells and total number $N_c$ of cancer cells in the system.
(b) Evolution of mean activity $u_i$ of immune system cells and mean activity $u_c$ of cancer cells.
System size $l_x=l_y=30$, $\Delta l=1$.
Initial condition: 
$20$ normal cells and $20$ immune system cells per box, $20$ cancer cells per box in a central discretized disc of diameter $D=5$, 
Gaussian distribution $P_u(u)$ of mean $\mu=0.5$ and standard deviation $\sigma=0.2$ for the initial activity of all cells.
Time step $\Delta t=1$,
rate constants $k_{ic}=0.01$, $k_{ci}=0.005$, $k_{cn}=5 \times 10^{-6}$, angle of interaction $\theta_{\rm inter}=\pi$, activity increase through
interaction $\epsilon=10^{-3}$,
cell speed $v=10^{-3}$, rate constant for velocity deflection $k_v=0.1$, maximum angle of velocity deflection $\theta_{\rm jump}=\pi/6$,
field associated with the thermostat $E= 10^{-4}$.
}
\label{3E}
\end{center}
\end{figure}
\begin{figure}[htb]
\begin{center}
\subfigure[]{\includegraphics[height=7cm]{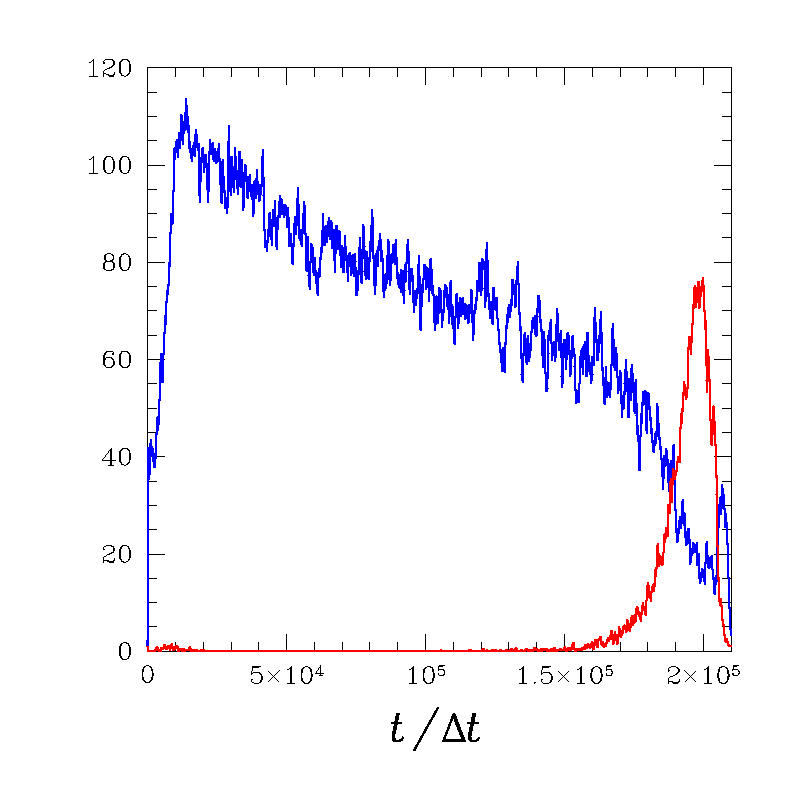}}
\subfigure[]{\includegraphics[height=7cm]{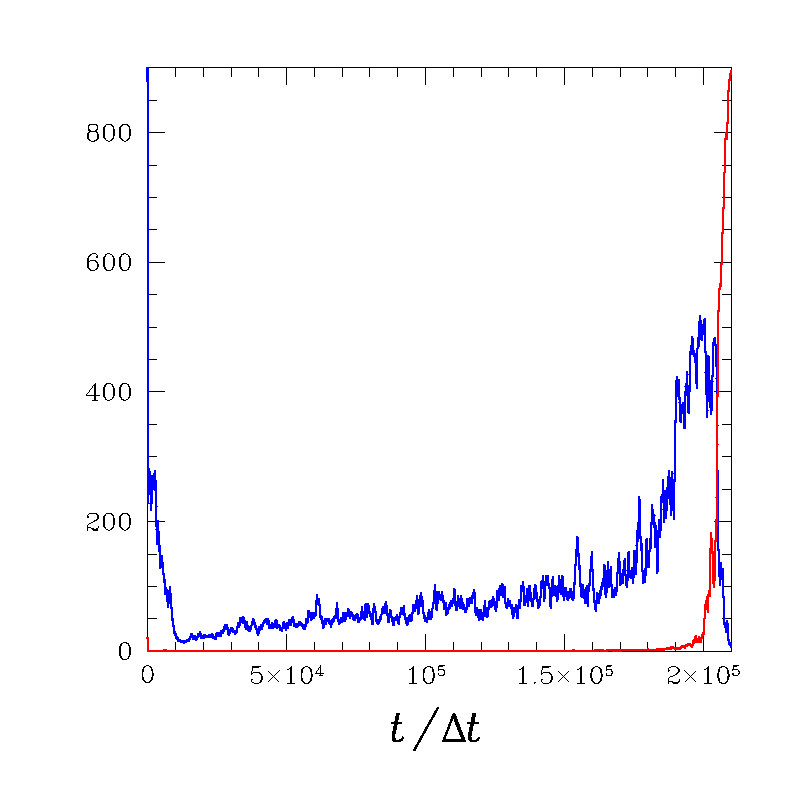}}
\caption{Evolution of (a) cluster number and (b) size of biggest cluster for cancer cells (red) and immune system cells (blue) averaged over a sliding window of $99$ time steps in the system shown in Fig. \ref{3E}.
}
\label{cluster3E}
\end{center}
\end{figure}

\subsection{The three Es of immunotherapy}
Figure \ref{3E} shows the evolution of the total number $N_c$ of cancer cells and the total number $N_i$ of immune system cells in the entire system.
A phenomenon analogous to the three Es observed in immunotherapy \cite{dunn2004} is reproduced by the 
inhomogeneous model for a critical value of the field $E$ depending on the values of the other parameters.
The number $N_c$ of cancer cells immediately drops during the nearly instantaneous elimination phase which is followed by a relatively small increase that was not observed in the homogeneous case \cite{jcp,physicaa}. Then, $N_c$ remains nearly constant during a long pseudo-equilibrium phase. 
Finally, $N_c$ explodes, becomes greater than the number of cells of the immune system, $N_i$, which in turn fluctuates with a large amplitude. 
Shortly after a very large fluctuation, $N_i$ cancels out and the cancer escapes from immunosurveillance.
As in the homogeneous case, this phenomenon is obtained for inefficient thermalization.
Cancer cells invade the entire system despite $k_{ci} < k_{ic}$: Inefficient control of activity fluctuations severely limits the chances of success of immunotherapy.  
Following the expression of the rate constant $\kappa_{ci}$ given in Eq. (\ref{kappacj}), the local important production of cancer cells is related to the learning of these cells, i.e. to the increase of their activity with respect to the neighboring immune system cells.
The learning process implemented in the model
consists of a small activity increase through cell interactions and a rate dependence on the relative activity of the interacting pairs.
This simple procedure is refined enough to reproduce the complex learning process of cancer cells, 
which can mislead the immune system and induce a harmful regulatory T cell response that wrongly orders a decrease in the production of immune cells.

The spatial description gives access to the evolution of the clusters of cancer cells and immune system cells. A cluster of cancer cell is defined as a continuous set of spatial boxes with neighboring boxes containing at least $20$ cancer cells, having a common side and not only a common corner. The value $20$ corresponds to the initial number of cancer cells per box in the small, localized tumor . An analogous definition is used for a cluster of
immune system cells. The size of a cluster is given by the number of boxes it contains. We use the Hoshen-Kopelman algorithm to build the clusters \cite{hoshen,nainville}.
Figure \ref{cluster3E} gives the evolutions of the averaged number of clusters and the size of the biggest cluster for each cell type for the same conditions as in Fig. \ref{3E}.

During the short transient initial period preceding the pseudo-equilibrium phase, the number of clusters of immune system cells rapidly increases while the size of the biggest cluster decreases from the total system size $30 \times 30$ to a small value around $10$, revealing the rapid fragmentation of the initial big cluster. During the same period, the number of clusters of cancer cells slightly fluctuates between $1$ and $0$ and the size of the biggest cluster instantaneously drops from $21$ to $0$, meaning that the number of cancer cells per box rapidly becomes smaller than $20$.
During the long pseudo-equilibrium phase, the number of clusters of immune system cells linearly decreases due to two phenomena, the coalescence of growing clusters and the decrease of the number of immune system cells, in agreement with the slight increase of the size of the biggest cluster.
During the same phase, the number of clusters and the size of the biggest cluster of cancer cells vanish, as if cancer has been eliminated.
Then, the number of clusters of cancer cells exponentially grows implying a global, nonlocal escape of cancer while the number of clusters of immune system cells drops. Eventually, the clusters of cancer cells coalesce and a single cluster invades the entire system while the immune system disappears: Cancer becomes out of control.
System size has an impact on the time at which an initially localized tumor invades the entire system but does not affect the qualitative behavior of the system.

\begin{figure}[htb]
\begin{center}
\includegraphics[height=12cm]{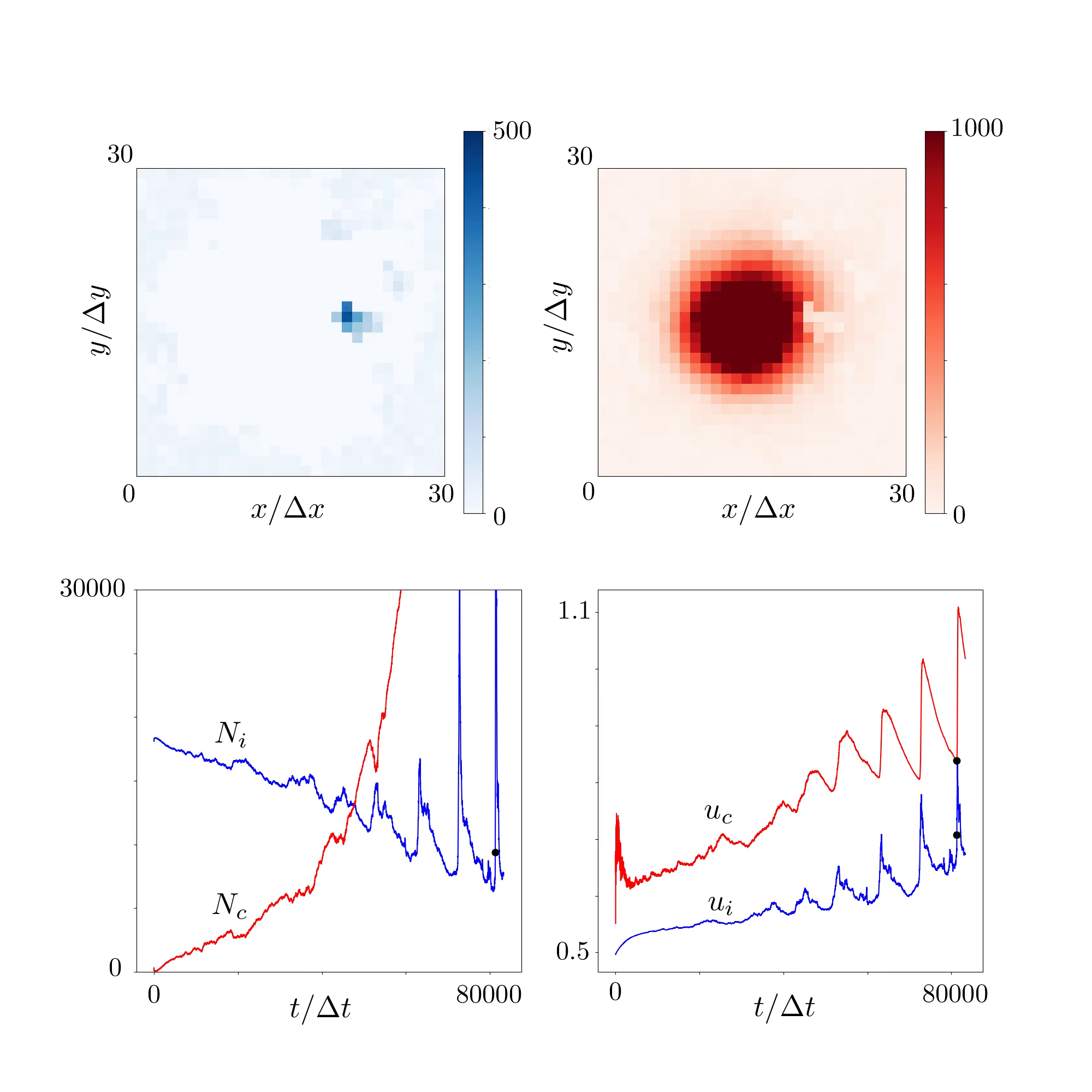}
\caption{Top left subfigure: Instantaneous spatial distribution of immune system cells (blue scale) and top right subfigure: Instantaneous 
spatial distribution of cancer cells (red scale) at time $t/\Delta t=81250$ indicated by a black disk in the bottom right subfigure and
associated with a minimum of mean activity $u_c$ for cancer cells.
Bottom left subfigure: Evolution of total number $N_i$ of immune system cells and total number $N_c$ of cancer cells in the system.
Bottom right subfigure: Evolution of mean activity $u_i$ of immune system cells and mean activity $u_c$ of cancer cells.
Same parameters as in the caption of Fig. 1 except the rate constant associated with the production of immune system cells, 
$k_{ic}=0.05$.
}
\label{start}
\end{center}
\end{figure}
\begin{figure}[htb]
\begin{center}
\includegraphics[height=12cm]{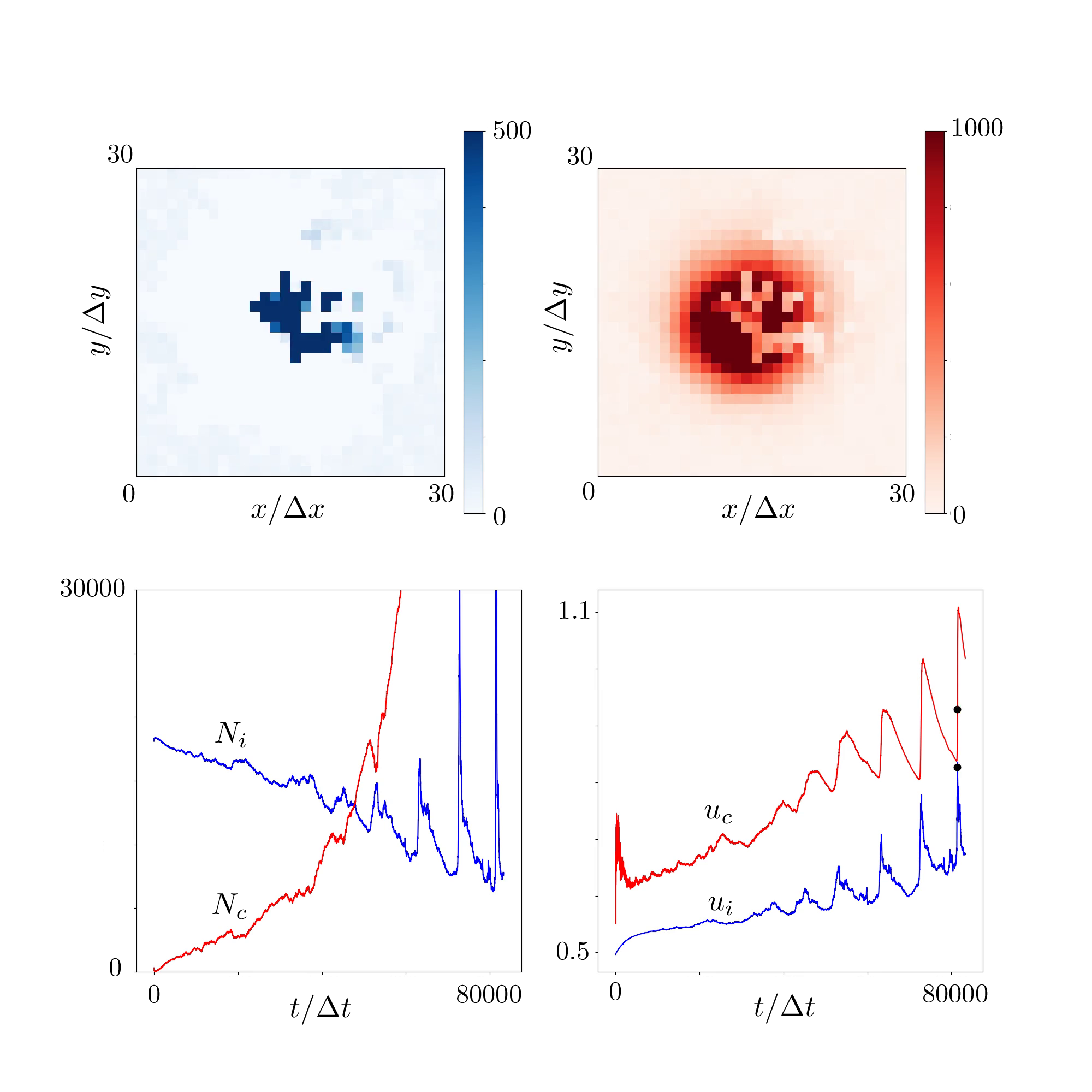}
\caption{Same caption as in Fig. \ref{start} at a different time $t/\Delta t=81935$ 
associated with a maximum of mean activity $u_i$ for immune system cells.
}
\label{mid}
\end{center}
\end{figure}

\begin{figure}[htb]
\begin{center}
\includegraphics[height=12cm]{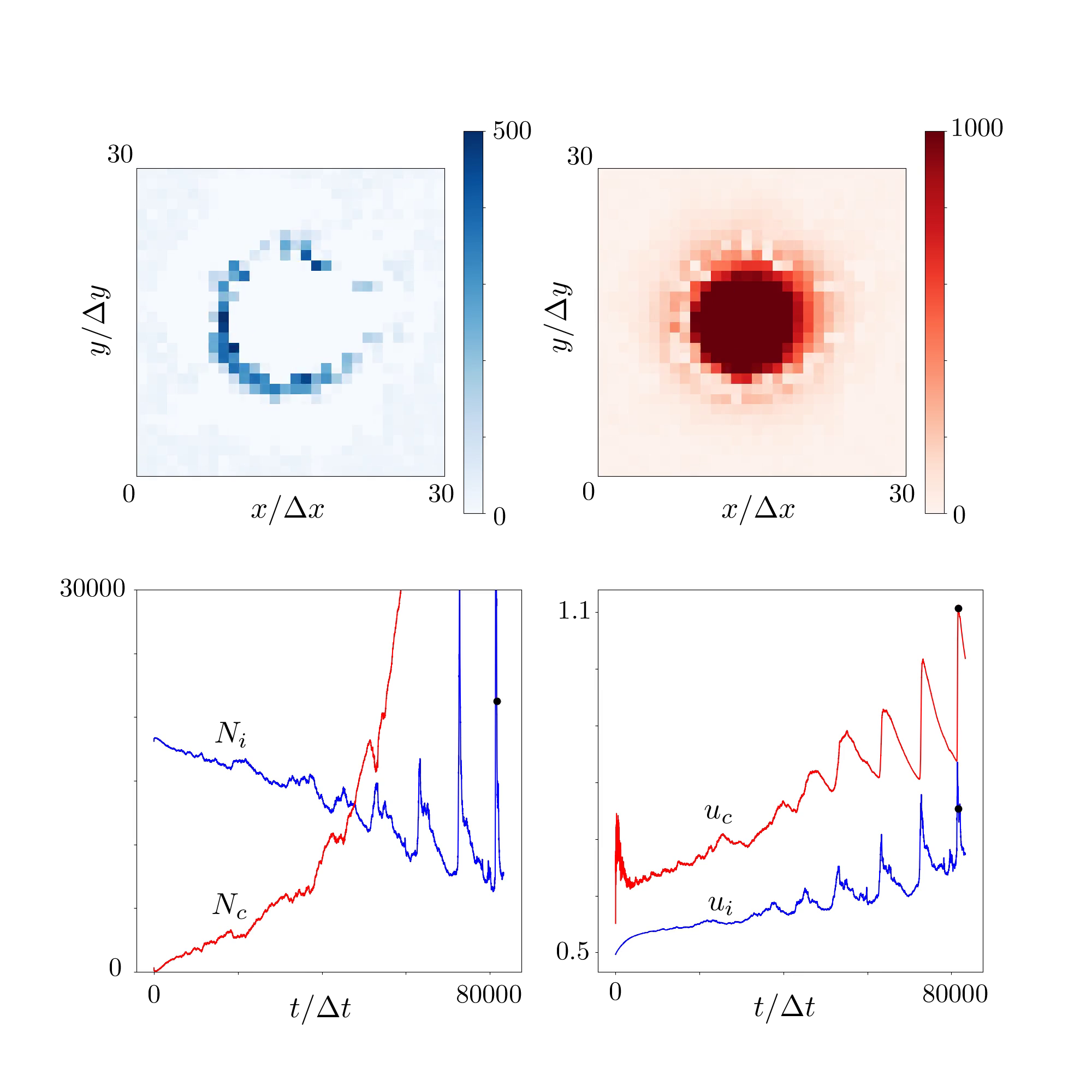}
\caption{Same caption as in Fig. \ref{start} at a different time $t/\Delta t=82040$ 
associated with a maximum of mean activity $u_c$ for cancer cells.
}
\label{end}
\end{center}
\end{figure}

\subsection{Oscillations of cell numbers and mean activities}
\begin{figure}[htb]
\begin{center}
\subfigure[]{\includegraphics[height=7cm]{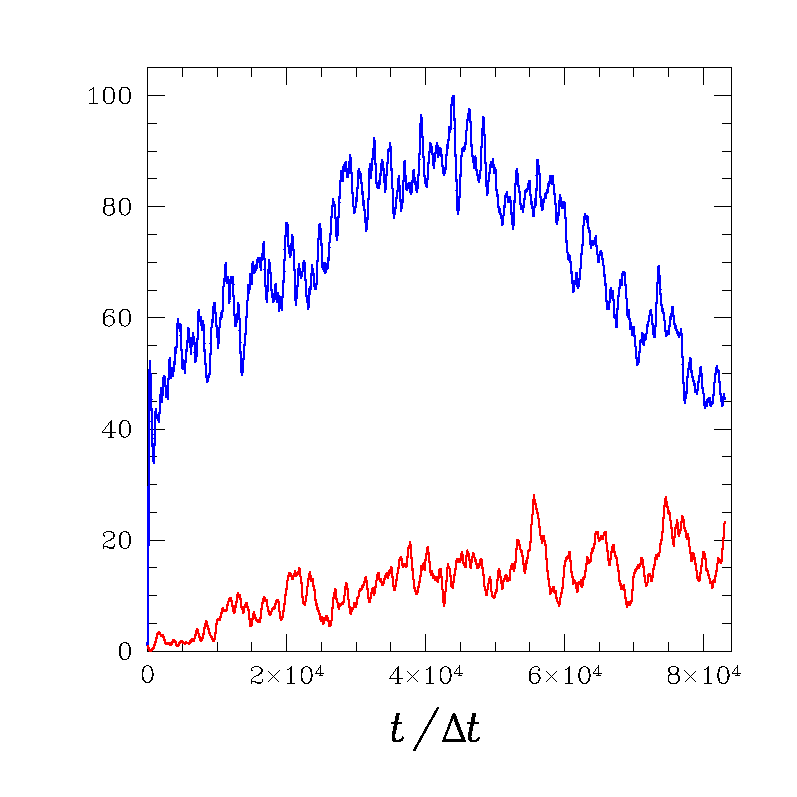}}
\subfigure[]{\includegraphics[height=7cm]{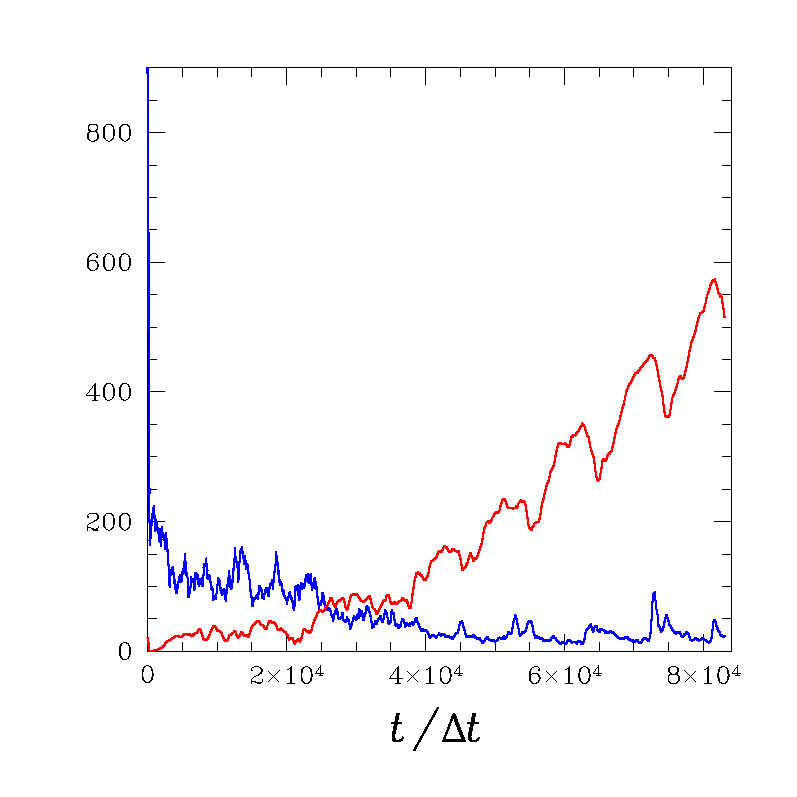}}
\caption{Evolution of (a) cluster number and (b) size of biggest cluster for cancer cells (red) and immune system cells (blue) averaged over a sliding window of $99$ time steps in the system shown in Figs. \ref{start}-\ref{end}.
}
\label{cluster}
\end{center}
\end{figure}

The main contribution of the inhomogeneous description is shown in Fig. \ref{start}. 
The total numbers $N_i$ of immune system cells and the mean activities, $u_c$, and $u_i$, of both types of cells display pseudo-oscillations of quite well-defined period and increasing amplitude. The total number $N_c$ of cancer cells rapidly becomes much larger than the total number $N_i$ of immune system cells and the chosen scale prevents from showing the end of the evolution of $N_c$.
The extrema of the mean activities $u_c$ and $u_i$ are associated with typical spatial configurations as shown in the different snapshots of
Figs. \ref{start}-\ref{end}.
A minimum of $u_c$ is rapidly followed by a maximum of $u_i$ and then a maximum of $u_c$.
As shown in Fig. \ref{start}, a minimum of $u_c$ corresponds to a nearly symmetric disc of cancer cells attacked by a small cluster of immune system cells due to a 
local favorable fluctuation of $u_i$.
Rapidly, the cluster of immune system cells grows and, after a small delay, the mean activity $u_i$ reaches a maximum, as displayed in Fig. \ref{cluster}. 
The cluster of immune system cells has an exotic shape which corresponds to a domain of the central tumor in which the number of cancer cells is depleted.
According to Fig. \ref{end}, the symmetry of the entire system is rapidly restored: due to their larger activities in average, the cancer cells deplete the immune system cells in the central part of the tumor, where they are more abundant. As a result, the immune system cells are pushed outwards the disc, forming an external ring.
Over a longer period of time, the ring of immune system cells will dissolve, leading to a minimum of $N_i$, a nearly homogeneous distribution of immune system cells, and a 
nearly perfect disc of cancer cells.
A new local fluctuation of $u_i$ will appear and trigger the cycle again.\\

As exemplified by the Brusselator model \cite{brusselator}, a chemical model presenting time oscillations contains the autocatalytic production
of one species $X$ to the detriment of another species $Y$ as well as a feedback mechanism which regenerates $Y$ and consumes $X$.
In the model of cancer/immune system interaction, the conditions on the relative activity of an interacting pair of cells 
given in Eqs. (\ref{ic}) and (\ref{ci}) introduce such a feedback mechanism.
In a small, localized cluster containing a high number of immune system cells with high activity, the autocatalytic production of immune cells by the process given in Eq. (\ref{ic}) is first favored. Then, the total number of cancer cells $N_c$ decreases. The effect on the average activity $u_i$ is non-trivial, the activity of the already present immune system cell being increased by $\epsilon$ and the newly formed immune system cell inheriting the lower activity of the interacting cancer cell. Simultaneously, the interaction tends to deplete the cancer cell population of its lowest activity elements. As a result, the number of cancer cells decreases but their mean activity $u_i$ increases. The autocatalytic production of cancer cells is then promoted by the process given in Eq. (\ref{ci}), which in turn tends to decrease the total number of immune cells $N_i$ but increase their average activity $u_i$. The system is again in the situation where the process given in Eq. (\ref{ic}) favors the autocatalytic production of immune cells and a new cycle begins.
Hence, activity-weighted rate constants for the autocatalytic productions of immune system cells and cancer cells is able to induce an oscillating behavior and
reproduce immune regulatory cycles observed in advanced cancer patients and revealed by oscillations of C-Reactive Protein concentration \cite{mahmoud,coventry}.

The interaction processes do not only select cells depending on their activity, they also introduce criteria on their velocity direction. 
Small values of the angle of interaction $\theta_{\rm inter}$ coupled to inefficient velocity randomization, i.e. small values
of both the rate constant $k_v$ and the maximum angle of velocity jump $\theta_{\rm jump}$, could induce a bias in the results for 
large values of $v/\Delta t $ with respect to box size $\Delta x$.
Specifically, a cluster of autocatalytically formed immune system cells with increased activity would move in a similar direction during a certain time, inducing a persistent anisotropy in cluster shape, which would induce a persistent deformation of tumor shape. Adapted parameter values could be chosen to simulate specific tumor shapes.
Conversely, efficient velocity randomization is essential to restore the initial tumor symmetry in the presence of velocity selection during cell interactions.
The simulation results shown in Figs. \ref{start}-\ref{end} are given for $\theta_{\rm inter}=\pi$, i.e. in the absence of selection of the angle of the interacting pair, which
prevents the formation of clusters with a selected velocity. 
Cell speed and time step obey $v=10^{-3}$ and $\Delta t =1$, respectively. The choice of $k_v=0.1$ for the rate constant of velocity deflection
leads to about 100 velocity jumps before a cell leaves a box, which consists of an efficient velocity randomization.  
Hence, a randomly occurring cluster of immune system cells surrounded by cancer cells gives rise to the isotropic propagation of a radial front of immune system cells.
Then, the cluster depletes from the center.
Indeed, the cancer cells in the boxes occupied by the initial cluster of immune system cells which have not been consumed
have an increased activity and are thus able to inhibit the immune system cells, leading to the formation of an external ring of immune system cells with radial velocity, similar to the one observed in Fig. \ref{mid}.\\

The evolution of the averaged number of clusters and size of the biggest cluster are given in Fig. \ref{cluster} for each cell type and for the parameter values of Figs. \ref{start}-\ref{end} leading to oscillations.
The number of clusters of cancer cells slightly increases during the entire duration of the simulation, the biggest cluster presenting a first fluctuating growing period followed by a faster increase accompanied by more regular oscillations.
The behavior of the size of the biggest cluster is close to the one of the mean number $N_c$ of cancer cells, which shows that the tumor swarms little, remains localized and grows around the initial cluster of cancer cells.
The size of the biggest cluster of cancer cells does not exceed $600$ boxes and remains smaller than system size equal to $30 \times 30$ cells,
making explicit that the oscillating behavior
shown in Figs. \ref{start}-\ref{end} is not affected by small size effects and boundary conditions. The dependence of the pseudo-period of oscillations on system size is 
checked in the next subsection.\\
The first increase of the number of clusters of immune system cells reveals the fragmentation of the big cluster initially occupying the entire system.
Then the number of clusters of immune system cells decreases while the size of the biggest cluster remains constant in average, proving that clusters disappear
due to immune system cells destruction and not to coalescence.
As shown in the snapshot of Fig. \ref{mid}, the maxima of $N_i$ correspond to the periodic appearance of a small cluster containing a large number of immune system cells in the central disc occupied by the large killing population of cancer cells, in which the immune system cells will be soon consumed before a new cycle.

\subsection{Sensitivity analysis}
Qualitatively different behaviors are observed as some relevant parameters of the model vary. 
In a homogeneous system, the field $E$ controlling activity fluctuations and one of the rate constants, $k_{ic}$ or $k_{ci}$, related to immune system cell or cancer cell production proved to have a significant impact on the phenomena \cite{physicaa}. 
In the case of an inhomogeneous system, we perform systematic analyses for increasing values of either $E$ or $k_{ic}$ with the other parameters set to the values shown in the caption of Fig. 1. \\

\begin{figure}[htb]
\begin{center}
\vspace{-4cm}
\includegraphics[height=14cm,angle=-90]{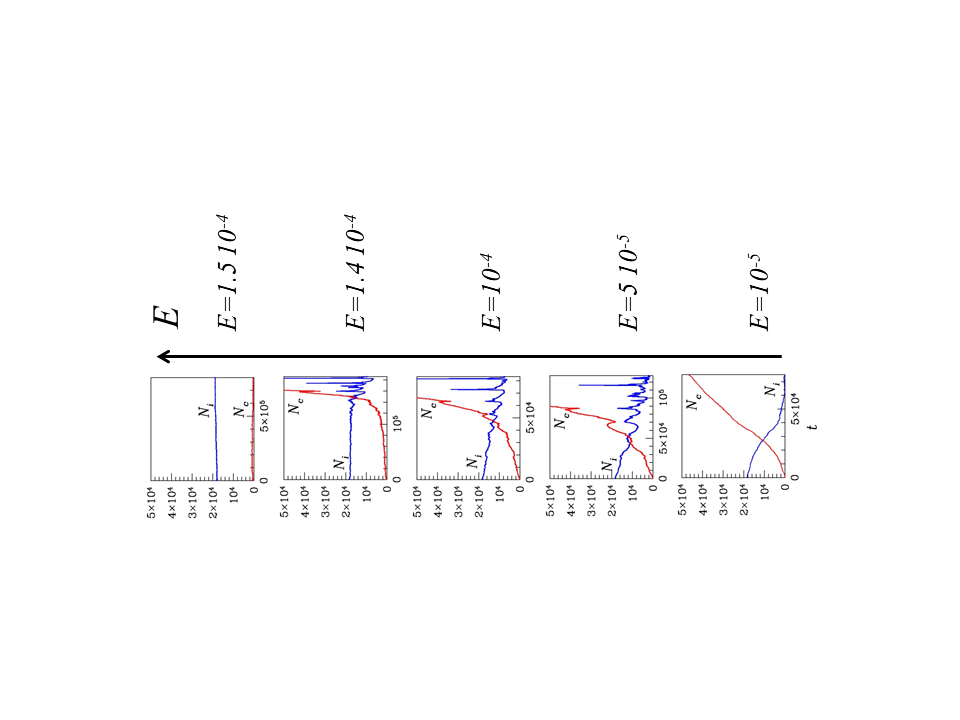}
\caption{The qualitatively different evolutions of the number $N_c$ of cancer cells (in red) and the number $N_i$ of immune system cells (in blue) observed as the field $E$ controlling activity fluctuations varies. The values of $E$ are indicated.
The constant associated with the production of immune system cells is set to $k_{ic}=0.05$. The other parameter values are given in the caption of Fig. 1.
}
\label{bifdiagNE}
\end{center}
\end{figure}

\begin{figure}[htb]
\begin{center}
\vspace{-4cm}
\includegraphics[height=14cm,angle=-90]{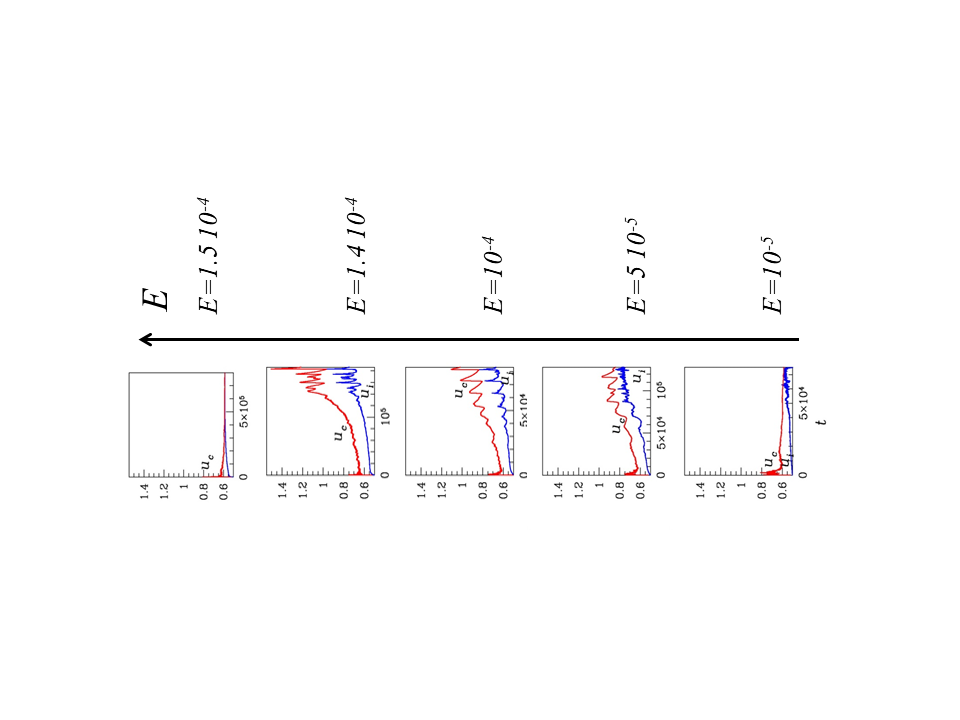}
\caption{Same caption as in Fig. \ref{bifdiagNE} for the evolutions of the activity $u_c$ of cancer cells (in red) and 
the activity $u_i$ of immune system cells (in blue) observed as the field $E$ varies ($k_{ic}=0.05$).
}
\label{bifdiagUE}
\end{center}
\end{figure}

Figures \ref{bifdiagNE} and \ref{bifdiagUE} illustrate typical evolutions of the numbers of cells and mean activities, respectively, as the field $E$ varies for 
$k_{ic}=0.05$, chosen smaller than $k_{ci}$. The same vertical scale is chosen for all subfigures of a given figure.
For a sufficiently small value of the field, typically $E \lesssim 10^{-5}$, corresponding to an inefficient control of activity fluctuations, a fast monotonous convergence toward elimination of the immune system and cancer growth is obtained. 
As $E$ increases, the evolutions of both cell numbers and activities become oscillating, in a chaotic manner for $E \simeq 5 \times 10^{-5}$ and a more regular
way illustrated in Figs. \ref{start}-\ref{end} for $E \simeq 10^{-4}$. For a critical value of the field, $E \simeq 1.4 \times 10^{-4}$, associated with 
a bifurcation, irregular fluctuations appear after a long induction period and the behavior illustrates the 3Es, with the final escape of cancer from immunosurveillance. For a slightly larger value of the field, $E \simeq 1.5 \times 10^{-4}$, the behavior is completely different and the system monotonically converges toward a steady state in which the activities of the two cell types are equal: Cancer is controlled, exactly as in the case of a sufficiently well thermalized homogeneous system \cite{physicaa}. The choice $k_{ic} > k_{ci}$ implies that the final steady number of immune system cells is larger than the final steady number of cancer cells.
\\

The results shown in Figs.  \ref{bifdiagNE} and  \ref{bifdiagUE} correspond to simulations stopped at different times 
such that the immune system cells have disappeared,
the number of cells in a spatial box exceeds a threshold set to $3000$ as obtained for the critical value $E \simeq 1.4 \times 10^{-4}$, 
or when the mean activities of the cancer cells or the immune system cells
are equal.
The two first cases obtained for  $E \leq 1.4 \times 10^{-4}$ correspond to cancer escape. The last case observed for $E \ge 1.5 \times 10^{-4}$ 
is associated with cancer control
and the simulation time is then supposed to mimic the life expectancy of a patient dying of a disease other than cancer.\\

\begin{figure}[htb]
\begin{center}
\includegraphics[height=7cm]{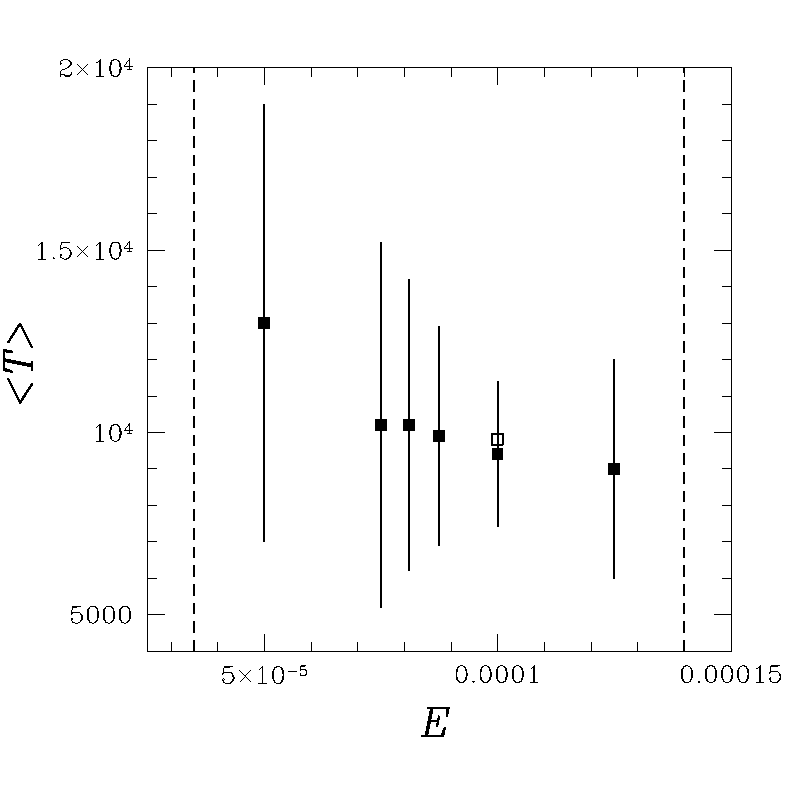}
\caption{Mean pseudo-period $\langle T \rangle$ of the oscillations versus field $E$ controlling activity fluctuations. The two dotted vertical lines limit the domain in which pseudo-oscillations are observed. The solid vertical bars give the uncertainty. 
The solid squares are obtained for a system of $30 \times 30$ boxes. The open square corresponds to a system of $20 \times 20$ boxes. 
The rate constant associated with the production of immune system cells is set to $k_{ic}=0.05$.
The other parameter values are given in the caption of Fig. 1.
}
\label{periodeT}
\end{center}
\end{figure}

Figure \ref{periodeT} represents the variation of the mean pseudo-period of oscillations of the total number of immune system cells $N_i$
versus $E$ in the field range where they are observed.
The uncertainty about their determination is minimum for $E \simeq 10^{-4}$ associated with the most regular oscillations of both $N_i$, $u_i$, and $u_c$ 
and increases as the boundaries of the domain of existence get closer.
Even for $E \simeq 10^{-4}$ the time between two maxima of $N_i$ decreases over the evolution, leading to rather large error bars for the pseudo-period.
The averaged pseudo-period $\langle T \rangle$ slightly decreases as $E$ increases. The value of $\langle T \rangle$ in a system of $30 \times 30$ boxes is satisfactorily compared to the result obtained in a system of $20 \times 20$ boxes. 
The difference between the two results is much smaller than the variability of the pseudo-period during time, 
proving that the simulation results deduced from a $30 \times 30$ system or even a $20 \times 20$ system are not sensitive to boundary effects and reliable.\\

\begin{figure}[htb]
\begin{center}
\vspace{-4cm}
\includegraphics[height=14cm,angle=-90]{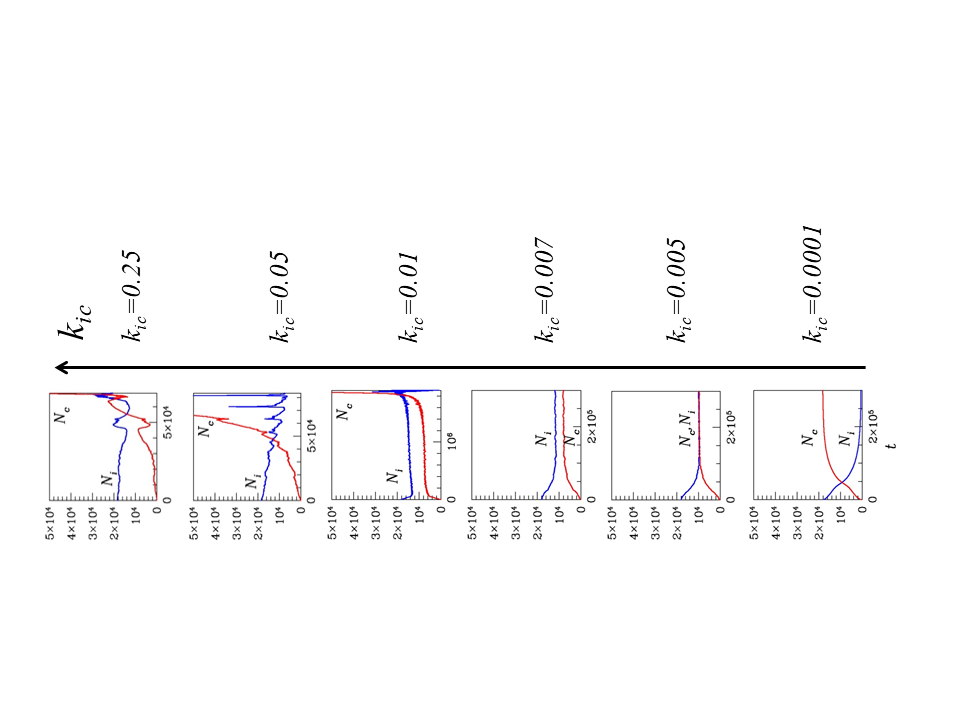}
\caption{The qualitatively different evolutions of the number $N_c$ of cancer cells and the number $N_i$ of immune system cells observed as the rate constant $k_{ic}$ controlling the autocatalytic production of immune system cells varies. The values of $k_{ic}$ are indicated. The other parameter values are given in the caption of Fig. 1 $(E=10^{-4})$.
}
\label{bifdiagNk}
\end{center}
\end{figure}

\begin{figure}[htb]
\begin{center}
\vspace{-4cm}
\includegraphics[height=14cm,angle=-90]{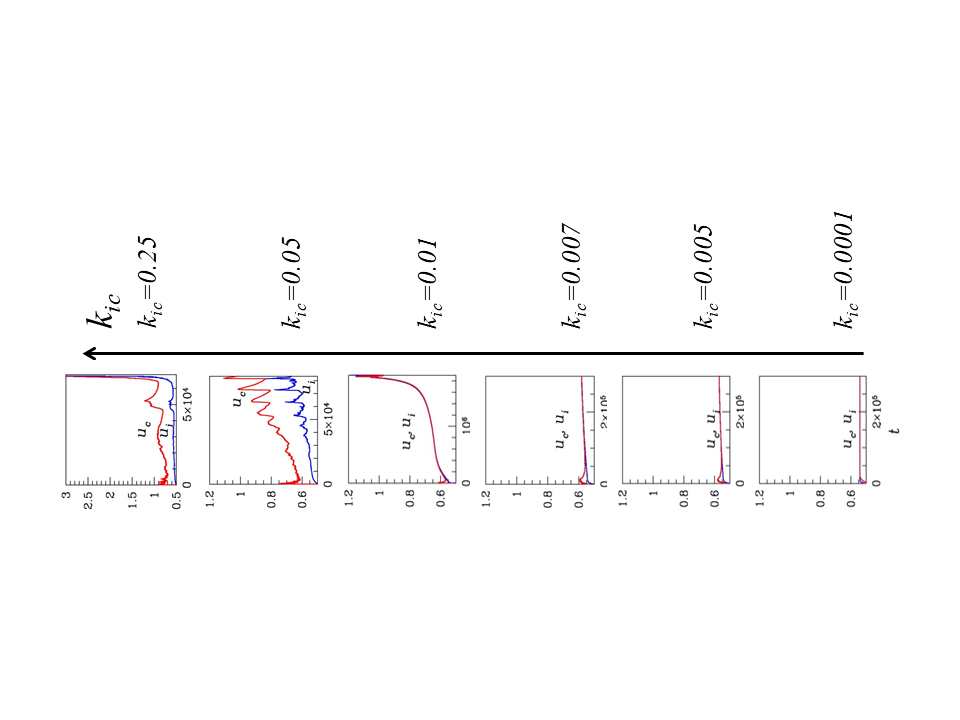}
\caption{Same caption as in Fig. \ref{bifdiagNk} showing different evolutions of the activity $u_c$ of cancer cells and the activity $u_i$ of immune system cells observed as the rate constant $k_{ic}$ varies $(E=10^{-4})$.
}
\label{bifdiagUk}
\end{center}
\end{figure}

The different typical behaviors encountered as the rate constant $k_{ic}$ controlling the autocatalytic formation of immune system cells increases are represented in Figs. \ref{bifdiagNk} and \ref{bifdiagUk}, all other parameters being fixed.

As shown in Fig. \ref{bifdiagUk} for sufficiently small values of $k_{ic}$ obeying $k_{ic} \lesssim 0.007$, the system rapidly and monotonously converges to a steady solution with the same final value $u_c^{\rm end}=u_i^{\rm end}$ of the mean activities of the two cell types $c$ and $i$.
The final cell numbers obey $N_c^{\rm end} > N_i^{\rm end}$ for $k_{ic} < k_{ci}$, they coincide for $k_{ic} = k_{ci}$, and obey
$N_c^{\rm end} < N_i^{\rm end}$ for $ k_{ci} < k_{ic} \lesssim 0.007$, as shown in Fig. \ref{bifdiagNk}. 
The symmetric roles played by the constants $k_{ic}$ and $k_{ci}$ clearly appear. In all these cases, cancer is controlled.
On the contrary, for sufficiently large $k_{ic}$, such that $k_{ic} \gtrsim 0.01$, cancer escapes from the surveillance of the immune system.
The case obtained for $k_{ic}=0.01$, already shown in Fig. \ref{3E}, illustrates the 3Es, with a long induction period and the eventual explosion of cancer cell number and both activities. For $k_{ic} \simeq 0.05$, rather regular oscillations of cell numbers and activities are obtained. Before disappearing for larger values of $k_{ic}$, the oscillations become erratic for $k_{ic} \simeq 0.25$. 
The value $k_{ic}=0.01$ can be considered as critical, smaller values leading to monotonous cancer control and similar mean activities of both cell types, larger values to oscillating cancer escape and larger mean activity for cancer cells. 
The results presented in Figs. \ref{bifdiagNk} and \ref{bifdiagUk} are nonintuitive since increasing $k_{ic}$, which should reinforce the production of immune system cells, induces the exit from the domain where cancer is controlled.
This result is explained by the feedback effect associated with the activity-dependence of the rate constant $\kappa_{ic}$ 
given in Eq. (\ref{kappaic}). At constant field $E$, increasing $k_{ic}$ locally depletes the population of cancer cells from its most active representatives, which induces a local increase of cancer cell activity. This fluctuation of cancer activity is then responsible for cancer cell production through the autocatalytic process given in Eq. (\ref{ci}), which explains the eventual explosion of the number of cancer cells and cancer escape from immunosurveillance.
In other words, a larger value of $k_{ic}$ leads to a larger critical value $E_c$ of the field associated with the 3Es and delimiting the
domains between cancer escape and cancer control: As shown in Figs. \ref{bifdiagNE} and \ref{bifdiagUE}, for $k_{ic}=0.05$
we find $E_c\simeq 1.4 \times 10^{-4}$ and in Figs. \ref{bifdiagNk} and \ref{bifdiagUk}, for $k_{ic}=0.01$ the bifurcation arises for a smaller critical field value $E_c\simeq 10^{-4}$.
A stronger stimulation of the immune system requires a more efficient regulation of activity fluctuations for cancer to be controlled.
\\

\begin{figure}[htb]
\begin{center}
\includegraphics[height=7cm]{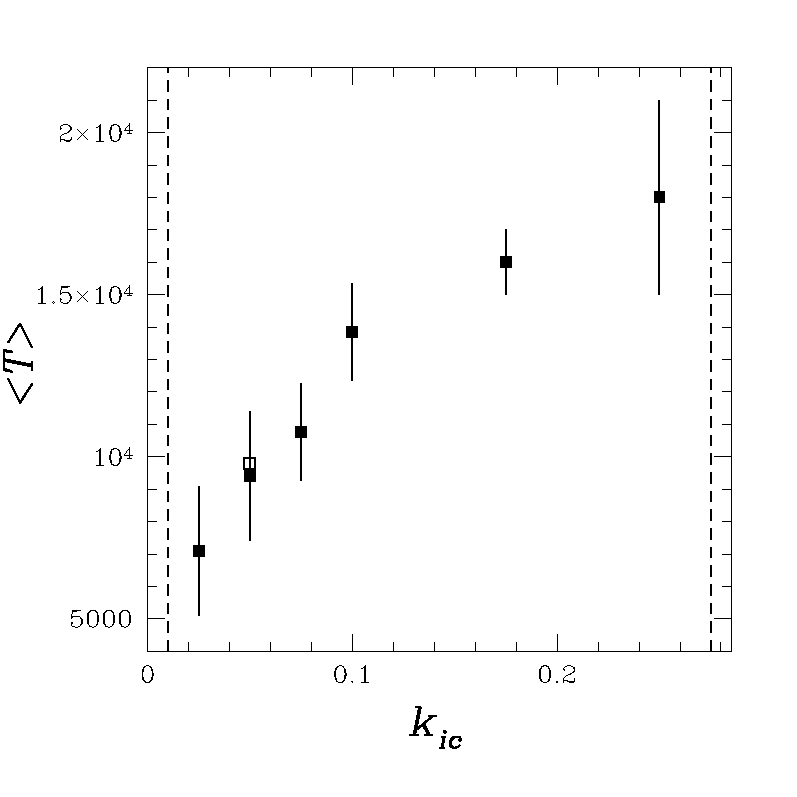}
\caption{Mean pseudo-period $\langle T \rangle$ of the oscillations versus rate constant $k_{ic}$ controlling autocatalytic production of immune system cells. The two dotted vertical lines limit the domain in which pseudo-oscillations are observed. The solid vertical bars give the uncertainty. The solid squares are obtained for a system of $30 \times 30$ boxes. The open square corresponds to a system of $20 \times 20$ boxes. The other parameter values are given in the caption of Fig. 1 $(E=10^{-4})$.
}
\label{periodeK}
\end{center}
\end{figure}

Figure \ref{periodeK} shows the variation of the mean pseudo-period of oscillations $\langle T \rangle$ versus  the rate constant $k_{ic}$ 
in the range in which they exist.
The uncertainty is much smaller than the variation of the averaged pseudo-period and we conclude without ambiguity that
 $\langle T \rangle$  increases as $k_{ic}$ increases. It was not easy to anticipate such a result, due, once again, to the antagonistic effects included in the model. At first glance, increasing $k_{ic}$ should lead to the faster production of immune system cells according to the autocatalytic process given in Eq. (\ref{ic}), which should decrease the time between two maxima of $N_i$ or $u_i$.
However, as shown in Figs. \ref{mid} and \ref{end}, the formation of a cluster of immune system cells associated with a maximum of mean activity $u_i$ is rapidly followed by a maximum for the mean activity $u_c$ of the cancer cells, due to the depletion of small activity cancer cells
induced by the process given in Eq. (\ref{ic}). Then, the very active cancer cells consume the immune system cells according to the autocatalytic process given in Eq. (\ref{ci}). As a result, the feedback effect on the activities is stronger than the increase of $k_{ic}$ and the time between two maxima of $N_i$ or $u_i$ is longer for larger $k_{ic}$.\\

We end the sensitivity analysis of the system to the field $E$ and the rate constant $k_{ic}$ with a few remarks about the effect of the other parameters. 
As explicitly shown in the case of a homogeneous system \cite{physicaa}, the role of $k_{ci}$ is symmetrical to the one of $k_{ic}$.
The results displayed in Figs. \ref{bifdiagNk} and \ref{bifdiagUk} for $k_{ic} < k_{ci}$, $k_{ic} = k_{ci}$, and $k_{ic} > k_{ci}$ illustrate this
property. 
A small change of cell speed $v$ modifies the critical values of the field and the constant $k_{ic}$ for which the 3Es are observed but does not qualitatively alter the behavior of the system.
However, another issue, far beyond the scope of this article, could be addressed in a different domain of parameter space. 
The model could be extended to the study of the anisotropic growth of a tumor by choosing smaller interaction angles $\theta_{\rm inter}$, larger cell speeds $v$, smaller rate constants $k_v$ of velocity randomisation and smaller deflection angles $\theta_{\rm jump}$ in order to favor the formation of clusters of cancer cells with a high activity moving in the same direction during a sufficiently long time to distort the tumor. 

\section{Conclusion}
The interactions between cancer and the immune system are complex and nontrivial antagonistic effects have been observed in patients with long-standing tumors and revealed by oscillations of some markers of inflammation \cite{coventry}.
We extend a model of cancer and immune system competition \cite{jcp,physicaa} to an inhomogeneous system in order to follow the dynamics of tumor growth at cell scale within the framework of thermostatted kinetic theory. The model includes cell-cell interactions mimicking mutation, death, division, regulation, and learning associated with
increased activity. Activity fluctuations are thermostatted to account for dissipation of information through nonspecific cell interactions.
We implement a kinetic Monte Carlo algorithm to directly simulate the kinetic equations in the spirit of DSMC method \cite{bird}. 
The impact of thermalization on the behavior of the system is important and a bifurcation occurs as the field controlling activity fluctuations
increases. For small field values, cancer invades the entire system whereas, for large field values, the system reaches a steady state 
with non vanishing numbers of cancer cells and immune system cells. Cancer is then controlled and the tumor reaches a stationary size.
For a critical value of the field  for which large activity fluctuations are eventually generated, 
the model reproduces the three Es of immunoediting \cite{dunn2004}, leading to the unexpected escape of cancer 
from immune system control after the elimination and equilibrium phases.
For a field value slightly smaller than the critical value, the model is sufficiently complex to reproduce an observed feature 
of tumor growth associated with rather regular oscillations of the immune response. 
We did not observe time oscillations in simulations of a homogeneous system \cite{jcp,physicaa}. Their stimulation requires non trivial initial conditions that 
are spontaneously generated in the simulations of the inhomogeneous system. In particular oscillations may begin when a cluster of immune system cells have a sufficiently higher activity than the neighboring cancer cells. The simulations have been performed for an efficient cell velocity randomization which ensures the rapid recovery of the initial symmetry of the tumor. Following the evolution of the mean number of clusters as well as the size of the biggest cluster reveals that the tumor remains a connected space apart from the loss of very small clusters on its periphery.
The model involves two autocatalytic reactions, the one producing cancer cells, the other, immune system cells, with a regulation through activity-dependent rate constants. 
Large-activity immune system cells kill low-activity cancer cells, leading to increased activity for the remaining cancer cells. These educated cancer cells are then able 
to kill low-activity immune system cells. Consequently, the mean activity of the remaining immune system cells increases and a new cycle may begin. 
It is worth noting that decreasing the rate constant $k_{ic}$ controlling the autocatalytic production of immune system cells has a similar effect
on the behavior of the system as increasing the field $E$. The same succession of bifurcations from cancer proliferation associated with
chaotic oscillations, periodic oscillations, long induction related to the 3Es, and cancer control is observed as shown when comparing
Figs. \ref{bifdiagNE} and  \ref{bifdiagNk}. Counterintuitively, larger rates of autocatalytic production of immune system cells require more efficient thermalization for cancer to be controlled, due to the feedback effect induced by activity-dependent rate constants. 
To our knowledge, the results of the model we developed are the first to account for the 3Es of immunotherapy and oscillations of immune system response 
within the framework of thermostatted kinetic theory and, in particular, at cell scale. As a perspective, macroscopic equations for concentrations and activities could be derived from the kinetic equations in the limit of appropriate space and time scaling \cite{carlo2,carlo3}. This derivation could give a microscopic interpretation at the scale of cell interactions to mean-field models, based on differential equations for macroscopic quantities, such as the concentration of different  
cell types \cite{ lejeune08,onofrio10,liu12,wilkie13,bi14,wenbo17,dritschel18,alvarez19}. \\

The results suggest improved treatment protocols, combining vaccination and chemotherapy at proper times. Reducing the dominance of T-regulatory cells  
over T-effector cells may be crucial in immunotherapy: Whereas T-effector cells kill cancer cells, T-regulatory cells are susceptible to improperly reduce the number of immune system cells, due to misleading information conveyed by cancer cells.
Vaccination can be used to maximize T-effector cell response and chemotherapy or radiotherapy, to deplete, {\it inter alia}, T-regulatory lymphocyte population.
Hence, the choice of the timing of vaccination, on the one hand, and chemotherapy, on the other hand, with respect to the immune regulatory cycle revealed for example by C-Reactive Protein oscillations, could be crucial in the treatment of some tumors.
The model, which reproduces intrinsic oscillations could be used to minimize the averaged number of cancer cells after periodic series of immune system boosting followed by immune system cell depletion at different times of the cycle. The results could give some hints on how a tumor reacts to periodic double disturbances mimicking vaccination and chemotherapy.

\section*{Conflict of interest}
All authors declare no conflicts of interest in this paper.

\end{document}